\documentclass[preprint,12pt]{elsarticle}

\usepackage{amssymb}
\usepackage{amsmath}

\journal{Annals of Physics}

\begin{document}

\begin{frontmatter}



\title{Solving a two-electron quantum dot model in terms of polynomial solutions of a Biconfluent Heun Equation}


\author[1,2]{F. Caruso}
\ead{francisco.caruso@gmail.com}

\author[2]{J. Martins} \author[2]{V. Oguri}
%\author[2]{V. Oguri}
\address[1]{Centro Brasileiro de Pesquisas F\'{\i}sicas - Rua Dr. Xavier Sigaud, 150, 22290-180, Urca, Rio de Janeiro, RJ, Brazil}
\address[2]{Instituto de F\'{\i}sica Armando Dias Tavares, Universidade do Estado do Rio de Janeiro - Rua S\~ao Francisco Xavier, 524, 20550-900, Maracan\~a, Rio de Janeiro, RJ, Brazil}
\begin{abstract}
The effects on the non-relativistic dynamics of a system compound by two electrons interacting by a Coulomb potential and with an external harmonic oscillator potential, confined to move in a two dimensional Euclidean space, are investigated. In particular, it is shown that it is possible to determine exactly and in a closed form a finite portion of the energy spectrum and the associated eigeinfunctions for the Schr\"{o}dinger equation describing the relative motion of the electrons, by putting it into the form of a biconfluent Heun equation. In the same framework, another set of solutions of this type can be straightforwardly obtained for the case when the two electrons are submitted also to an external constant magnetic field.

\end{abstract}


\end{frontmatter}

% \linenumbers

\section{Introduction}
\label{int}

Let us consider the problem of two interacting electrons confined to move in an external
harmonic oscillator potential of frequency $\Omega$. As an example of such system one can quote electrons confined in a nanometer-scale semiconductor
structure, often called a quantum dot~\cite{Reimann}, where the electrons have been shown to exhibit a two dimensional behavior~\cite{Sikorski}. Previous numerical calculations suggest that the harmonic oscillator potential can be successfully employed to describe two-electron quantum dots~\cite{Merkt}.
Thus, this kind of system can be described by the time independent Schr\"{o}dinger equation, in atomic units ($\hbar = m = e =1$),
\begin{equation}\label{original_equation}
\left\{ -\frac{1}{2} (\nabla_1^2 + \nabla_2^2) + \frac{\Omega^2}{2} (r_1^2 + r_2^2) + \frac{1}{|\vec r_1 - \vec r_2|}  \right\} \Psi(\vec r_1, \vec r_2) = E_T\, \Psi(\vec r_1, \vec r_2)
\end{equation}
where the subscripts 1 and 2 refer to each one of the electrons. The $\vec r_i$ are two-dimensional vectors with length $r_i = |\vec r_i|$. Introducing the usual relative and center of mass coordinates, $\vec r = \vec r_1 - \vec r_2$ and $\vec R = (\vec r_1 + \vec r_2)/2$, Eq.~(\ref{original_equation}), with the choice
$\Psi(\vec r_1, \vec r_2) = \chi(\vec R) \psi(\vec r)$, gives rise to the following pair of equations:
\begin{eqnarray}\label{two_equations}
&&\left[ -\frac{1}{2} \nabla_{\vec R}^2 + \frac{\omega_R^2}{2} R^2 \right] \chi(\vec R) = 2 \epsilon\, \chi(\vec R)\\
\nonumber \ \\
&&\left[ -\frac{1}{2} \nabla_{\vec r}^2 + \frac{\omega^2}{2} r^2 + \frac{1}{2r} \right] \psi(\vec r) = \frac{1}{2} \eta\, \psi(\vec r)
\end{eqnarray}
where one has defined the frequencies $\omega_R=2\Omega$ and $\omega = \Omega/2$. The total energy is given by
$$ E_T = {\epsilon} + \eta$$

The equation for the center of mass coordinate is the usual non relativistic equation for the harmonic oscillator in two dimensions, which means that the energy eigenvalues are
$$\epsilon = \omega_R (n_R +1)$$
with $n_R = 0,1,2,3,\cdots$. The 2$D$ radial Schr\"{o}dinger equation for the relative coordinate can be obtained by introducing the polar coordinates $(r,\theta)$ and putting its solution in the form
\begin{equation}
\label{main_sol}
\psi(\vec r) = r^{-1/2}\, u(r) e^{\pm i l \theta} = R(r) e^{\pm i l \theta}
\end{equation}
where $l$ is the integer angular momentum quantum number of the two-dimensional system. The radial function $u(r)$ should satisfy the following equation:
\begin{equation}\label{radial_equation}
\frac{\mbox{d}^2u(r)}{\mbox{d}r^2} + \left[ 2\eta - \frac{1}{r} - \omega^2 r^2 - \frac{(l^2-1/4)}{r^2} \right] u(r) = 0
\end{equation}

In Ref.~\cite{Taut}, where the problem of two electrons in an external oscillator potential is studied in three dimensions, it is shown that the above radial equation is \textit{quasi-exactly solvable}~\cite{Turbiner_1, Turbiner_2, Usheridze}, which means that it is possible to find exact simple solutions for some, but not all, eigenfunctions, corresponding to a certain infinite set of discrete oscillator frequencies. In Refs.~\cite{Taut_94,Taut_2000,Taut_2010}, the same problem was considered in the presence of a homogeneous magnetic field $B$. In this case, the Schr\"{o}dinger equation can be written, in the notation of Ref.~\cite{Taut_2010}, as
\begin{equation}\label{radial_equation_B}
\left\{-\frac{1}{2}\frac{\mbox{d}^2}{\mbox{d}r^2} + \frac{1}{2} \frac{(m^2-1/4)}{r^2} + \frac{1}{2}\widetilde{\omega_r}^2 r^2 + \frac{1}{2r}  \right\} u(r) = \widetilde{\epsilon_r} u(r)
\end{equation}
where $\widetilde{\omega_r} = \widetilde{\omega}/2$, $\widetilde{\epsilon_r} = \epsilon_r - m\omega_c/4$, with $\omega_c = B/c$ and $\widetilde{\omega} = \sqrt{\omega_0^2 + (\omega_c/2)^2}$; $\omega_0$ is the frequency of the harmonic confinement potential and $\omega_c$ is the cyclotron frequency of the magnetic field. It is straightforward to see that, putting $m=l$, Eq.~(\ref{radial_equation_B}) is the same as our Eq.~(\ref{radial_equation}) and thus it is sufficient to discuss here the simplest case of  Eq.~(\ref{radial_equation}).

The main scope of this paper is to propose a different theoretical approach to find analytical solutions for the two-dimensional problem of quantum dots that, we hope, can be applied to other simple two-body physical systems and, at the same time, provide us a way to confirm or not the previous results \cite{Taut_94,Taut_2000,Taut_2010}. The starting point is to transform Eq.~(\ref{radial_equation}) into a particular Heun's differential equation, since it is well known that the solution for some quantum-mechanical systems described by a radial Schr\"{o}dinger equation can be obtained in terms of those of the Biconfluent Heun Equations (BHE) \cite{Lemieux}. The Eq.~(\ref{radial_equation}) is indeed a particular case of a Schr\"{o}dinger equation for a kind of confining potential energy given by
\begin{equation}\label{conf_potential}
U(r) = \pm \frac{a}{r} + br + cr^2, \qquad c>0
\end{equation}
which was formally studied in the literature \cite{Chaudhuri_1, Chaudhuri_2, Arriola}.

We cannot neglect a priori the possibility of finding other particular solutions for the quantum dot states within this new technique. 

\section{The Biconfluent Heun's Equation}
\label{former_prediction}

The radial Schr\"{o}dinger equation for the confinement potential (with a repulsive Coulombian potential between the two electrons), given by Eq.~(\ref{conf_potential}), can be written, in atomic units, as
\begin{equation}\label{conf_equation}
u^{\prime\prime} + \left[ 2\left(E - \frac{a}{r} - b r - c r^2\right) - \frac{\ell(\ell+1)}{r^2} \right] u(r) = 0
\end{equation}
and can be straightforward transformed into the canonical form of BHE~\cite{Ronveaux}, namely,
 \begin{equation}\label{bhe}
xy^{\prime\prime} + (1 + \alpha - \beta x - 2x^2) y^\prime + \left[(\gamma - \alpha -2)x - \frac{1}{2} [\delta + (1+\alpha) \beta]\right] y(x) = 0
\end{equation}

A simple comparison between Eqs.~(\ref{radial_equation}) and (\ref{conf_equation}) allow to define the parameters:
$$ E = \eta; \quad 2c = \omega^2; \quad
  a = \frac{1}{2}; \quad b = 0; \quad \ell = \pm l - 1/2
$$

Now, putting~\cite{Leaute}
\begin{equation}
\label{u_into_y}
u(r) = r^{\ell+1} e^{-(\beta_F r+ \frac{1}{2} \alpha_F r^2)}\, y(r)
\end{equation}
where $x = \sqrt{\alpha_F}\, r$ and
$$\alpha_F = \sqrt{2c} = \omega; \quad \beta_F =  \sqrt{\frac{2}{c}}\,b = 0; \quad \epsilon_F= \beta_F^2 + 2E = 2 \eta, $$
it follows that Eq.~(\ref{conf_equation}) becomes
\begin{eqnarray}
 \nonumber x y^{\prime\prime} &+& \displaystyle \left[ 2(\ell+1) \sqrt{\alpha_F} - \frac{2\beta_F}{\sqrt{\alpha_F}} x - 2 x^2 \right] y^\prime + \\
\nonumber \ \\
   &+& \displaystyle \left\{\left[ \frac{\epsilon_F}{\alpha_F} - (2\ell +3)\right]x + \frac{a-2\beta_F(\ell+1)}{\sqrt{\alpha_F}}\right\} y(x) = 0
\end{eqnarray}

This is exactly Eq.~(\ref{bhe}) provide one defines the four Heun parameters as
$$\alpha = 2(\ell+1)\sqrt{\alpha_F} -1; \quad \beta= \frac{2\beta_F}{\sqrt{\alpha_F}}; \quad \gamma = \frac{\epsilon_F}{\alpha_F} + 2(\ell+1) (\sqrt{\alpha_F} -1)$$
and
$$ \delta = \frac{2}{\sqrt{\alpha_F}}\, \left[-a + 2\beta_F (\ell+1)(1- \sqrt{\alpha_F}) \right]$$

Turning back to the original parameters ($\omega$, $\eta$, $l$) (and keeping the choice $\ell + 1 = l +1/2$, which guarantees that the radial part of the wave function $\psi$ will be regular at $r=0$, for all values of $l$), we get
$$\alpha = (2l+1)\sqrt{\omega} - 1; \quad \beta =0; \quad \gamma = \frac{2\eta}{\omega} + (2l+1) (\sqrt{\omega} -1); \quad \delta = - \frac{1}{\sqrt{\omega}}$$
and, therefore, solving Eq.~(\ref{radial_equation}) is equivalent to solve the equation
\begin{equation}
\label{partic_bhe}
x\, y^{\prime\prime}(x) + [1 + \alpha - 2x^2]\, y^\prime + [-\delta/2 + (\gamma - \alpha -2)x]\,y(x) = 0
\end{equation}

\section{Polynomial solutions}\label{poly}

The polynomial solutions of Eq.~(\ref{partic_bhe}), a particular case of Eq.~(\ref{bhe}), have been analysed by many authors~\cite{Hautot-1, Hautot-2, Rovder, Urwin}.
When $\alpha$ is not a negative integer, one can denote by $N(\alpha,\beta,\gamma,\delta;x)$ the series solution that can be written as
\begin{equation}
\label{definition_N}
y_1 (x) = N(\alpha,\beta,\gamma,\delta;x) = \sum_{p=0}^{\infty} \frac{A_p}{(1+\alpha)_p}\, \frac{x^p}{p!}
\end{equation}
with
$$(\alpha)_p = \frac{\Gamma(\alpha +p)}{\Gamma(\alpha)}, \quad p\geq 0$$

The two first coefficients are $A_0 =1$ and $A_1 = [\delta+ \beta(1+\alpha)]/2 = \delta/2$ (for $\beta=0$). The substituting of $y_1(x)$, given by Eq.~(\ref{definition_N}), as well as its first and second derivatives into Eq.~(\ref{partic_bhe}) leads to
\begin{eqnarray*}
(\alpha +1) \left[A_1 - \frac{\delta}{2} A_0\right] &+& \left[A_2 - \frac{\delta}{2} A_1 + (\gamma - \alpha -2) (\alpha+1) A_0 \right] x +\\
\ \\
 &+& \sum_{p=1}^\infty \frac{x^{p+1}}{(p+1)!(\alpha+1)_{p+1}} \, \left[A_{p+2} - \frac{\delta}{2} A_{p+1} + \right.\\
 \ \\
 &+& \left.(\gamma -\alpha-2-2p) (p+1) (p+\alpha+1)A_p \right] = 0
\end{eqnarray*}
Therefore, defining $\delta^\prime \equiv - \delta/2$, one should have
\begin{eqnarray*}\label{recur}
  \ &\,& \delta^\prime A_0 + A_1 = 0 \\
  \ \\
  \ &\,& (\gamma -\alpha -2) (\alpha +1) A_0 + \delta^\prime A_1 + A_2 = 0\\
  \ \\
  \ &\,&  (\gamma -\alpha-2-2p) (p+1) (p+\alpha+1)A_p + \delta^\prime A_{p+1} + A_{p+2} = 0
\end{eqnarray*}
or, in matrix notation,

$$\begin{pmatrix}
  \delta^\prime &1              &0              &0              &\ldots       &\ldots           &0 \\
  \gamma_1       &\delta^\prime &1              &0              &\ldots       &\ldots           &0 \\
  0              &\gamma_2       &\delta^\prime &1              &0            &\ldots           &0 \\
  0              &0              &\gamma_3       &\delta^\prime &1            &\ldots           &0 \\
  \vdots         &\vdots         &0              &\ddots         &\ddots       &\ddots           &\ldots \\
  \vdots         &\vdots         &\vdots         &\vdots         &\gamma_{p-1} &\delta^\prime   &1 \\
  0              &0              &0              &0              &0            &\gamma_p         &\delta^\prime \\
\end{pmatrix}
\begin{pmatrix}
  A_0 \\
  A_1 \\
  A_2 \\
  \vdots \\
  \vdots \\
  A_{m-1} \\
  A_m \\
\end{pmatrix}
= 0
 $$

From the above recurrence expressions, the function $N$, defined in Eq.~\ref{definition_N}, becomes a polynomial of degree $n$ if and only if
$$\gamma - \alpha - 2 = 2n; \qquad n=0,1,2,3,... \qquad \mbox{and} \qquad A_{n+1}=0$$
and so the $\gamma_p$ factors of the determinant are given by
$$\gamma_p = 2 (n-p)(p+1)(p+1+\alpha), \qquad p\geq 2$$

In terms of the original parameters of the problem, this relation gives rise to a quantum condition between the energy $\eta$ and the frequency $\omega$, since
\begin{equation}
\label{quantum_rel}
\gamma - \alpha - 2 = \left( 2\frac{\eta}{\omega} -2l - 2 \right)= 2n \quad \Rightarrow \quad \eta_{nl} = \left(n + l + 1\right)\, \omega
\end{equation}

Note that in the case of the polynomial solutions, although they are not the most general one, it is straightforward to find the relation between the energy spectrum and a given oscillator frequency $\Omega = 2 \omega$, diversely form the approach followed in Ref.~\cite{Taut}.

 Each coefficient $A_{n+1}$ is a polynomial of degree $n+1$ in $\delta^\prime$ and there are at most $n+1$ suitable values of $\delta^\prime$~\cite{Ronveaux}.

Thus, when $p=n$, the $n$ roots can be determined by solving the $p\times p$ determinant $\Delta$ of the matrix $\mathbb{M}$ defined below:

$$
\Delta = \det \mathbb{M} =
\begin{Vmatrix}
  & \delta^\prime   & 1              & 0                   & \ldots         & \ldots & \ldots            & 0\\
  & \gamma_1         & \delta^\prime & 1                   & \ldots         & \ldots & \ldots            & 0 \\
  & 0                & \gamma_2       &\delta^\prime       & 1              & \ldots & \ldots            & 0 \\
  & \vdots           & 0              &\gamma_3             &\delta^\prime & 1      & \ldots            & 0 \\
  & \vdots           & \vdots         & 0                   & \ddots         & \ddots & \ddots            & \vdots\\
  & \vdots           & \vdots         & \vdots              & \vdots         & \ddots & \delta^\prime    & 1 \\
  & 0                & 0              & 0                   & 0              & 0      & \gamma_{n-1}      &\delta^\prime \\
\end{Vmatrix}
=0
$$

This is the determinant of a tridiagonal matrix and can be calculated by the decomposition $\mathbb{M} = \mathbb{L} \mathbb{V}$, such that
\begin{equation}
\label{leu}
\det \mathbb{L} =
\begin{Vmatrix}
  & 1          & 0             & 0         & \ldots    & 0\\
  & \ell_2     & 1             & 0         & \ldots    & 0 \\
  & 0          & \ell_3       &\ddots      & \ddots    & 0 \\
  & 0          & 0            &\ddots      & \ddots    & 0 \\
  & 0          & 0            &0           & \ell_n    & 1 \\
\end{Vmatrix}; \
\det \mathbb{V} =
\begin{Vmatrix}
  & v_1        & c_1          & 0         & \ldots    & 0\\
  & 0          & v_2          & c_2       & \ldots    & 0 \\
  & 0          & 0            &\ddots     & \ddots    & 0 \\
  & 0          & 0            &\ddots     & v_{n-1}   & c_{n-1} \\
  & 0          & 0            &0          & 0         & v_n \\
\end{Vmatrix}
\end{equation}
where
\begin{eqnarray*}
v_1 &=& \delta^\prime \\
c_1 &=& c_2 = \ldots = c_{n-1} = 1 \\
\ell_i &=& \frac{\gamma_{i-1}}{v_{i-1}} \\
v_i &=& \delta^\prime - \ell_i =  \left(\delta^\prime - \frac{\gamma_{i-1}}{v_{i-1}}\right), \quad i=2,\ldots n
\end{eqnarray*}

Since both matrixes in Eq.~(\ref{leu}) are triangular matrixes, each determinant is equal to the product of the elements of the principal diagonal. Therefore,
\begin{equation}
\label{determ}
\det \mathbb{M} = \det \mathbb{L}\cdot \det \mathbb{V} = 1 \cdot v_1 \cdot v_2 \ldots v_{n-1}\cdot v_n
\end{equation}
or
\begin{eqnarray*}
  \det\mathbb{M} &=& \delta^\prime \times \left[\delta^\prime - \frac{\gamma_1}{\delta^\prime} \right] \times \left[\delta^\prime -\frac{\gamma_2}{\displaystyle \delta^\prime - \frac{\displaystyle \gamma_1}{\delta^\prime}} \right]\times \left[\delta^\prime -\frac{\gamma_3}{\displaystyle \delta^\prime - \frac{\gamma_2}{\displaystyle \delta^\prime - \frac{\displaystyle\gamma_1}{\delta^\prime}}} \right] \times \\
  \, &\times & \ldots \times  \left[\delta^\prime -\frac{\gamma_{n-1}}{\displaystyle \delta^\prime - \frac{\gamma_{n-2}}{\displaystyle \delta^\prime - \frac{\ddots}{\displaystyle\frac{\ddots}{\displaystyle \delta^\prime - \frac{\gamma_1}{\delta^\prime}}}}} \right]
\end{eqnarray*}

Let us now define $d_1 \equiv \delta^\prime$ and
\begin{eqnarray*}
  d_2 &=& \delta^\prime d_1 - \gamma_1 \\
  d_3 &=& \delta^\prime d_2 - \gamma_2 d_1  \\
  d_4 &=& \delta^\prime d_3 - \gamma_3 d_2  \\
  \vdots &=& \vdots \\
  d_{n} &=& \delta^\prime d_{n-1} - \gamma_{n-1}d_{n-2} \quad \mbox{for} \quad n\geq2.
\end{eqnarray*}
Thus, it follows from Eq.~(\ref{determ}) that
$$\det \mathbb{M} = d_1 \times \left(\frac{d_2}{d_1}\right) \times \left(\frac{d_3}{d_2}\right) \times \left(\frac{d_4}{d_3}\right) \times \ldots \times \left( \frac{d_{n-1}}{d_{n-2}}\right) \times \left( \frac{d_{n}}{d_{n-1}}\right)$$

\noindent or, simply, for each $n\times n$ matrix $\mathbb{M}$,
\begin{equation}
\label{det_value}
\det \mathbb{M} = d_{n}
\end{equation}

The condition $\det \mathbb{M} = 0$ will give rise to a polynomial in $\delta^\prime$ and its independent roots will be denoted by $\delta^\prime_{\mu}$ $(0 \leq \mu \leq n)$. From this procedure one gets the roots of $1/\sqrt{\omega}$ for which the physical polynomial solutions can be constructed.
Once these roots are determined, the explicit polynomial form that depends on two quantum numbers ($n,l)$ can be written as:
\begin{equation}
\label{ynl}
y_{nl} (x) = N(\alpha,0,\alpha+2(n+1),\delta^\prime_{n};x) = \sum_{p=0}^{n} \frac{A_p}{(1+\alpha)_p}\, \frac{x^p}{p!}
\end{equation}
remembering that $A_0=1$, $A_1=-\delta^\prime$ and, for $p\geq 2$, $A_p$ is a polynomial in $\delta^\prime$ and $\alpha = (2l+1) \sqrt{\omega} -1$.

%When the root $\delta^\prime_{\mu} = 0$, the wave function can be written in terms of the confluent hypergeometric function ${}_1F_1(a;c;x) = \Phi(a;c;x)$, as~\cite{Ronveaux}
%\begin{equation}
%\label{particular-case}
%y_{nl} (x) = N(\alpha,0,\gamma,0;x) = \Phi \left( \frac{1}{4} (\alpha + 2 -\gamma); 1 + \frac{\alpha}{2}; x^2 \right)
%\end{equation}
%remembering that $(\alpha + 2 -\gamma)/4 = -n/2$ and $\alpha = (2l+1)\sqrt{\omega} - 1$.

\section{Numerical results}\label{results}

Let us now proceed to the explicit determination of the wave functions and respective energy levels for the ground state and other few lower states, namely $n=1,2,\ldots, 5$.

The first step is to determinate the roots of $1/\sqrt{\omega}$ for each values of $n$ and $l$. Some values corresponding to $l=0$ states are given in Table~\ref{roots}.

\renewcommand{\arraystretch}{1.8}
\begin{table}[hbtp]
\caption{\small Some roots of $1/\sqrt{\omega}$ for $l =0$ and different values of $n$, compared to those found in Ref.~\cite{Taut}.}
\vspace*{0.2cm}
 \label{roots}
\centerline{\small
\begin{tabular}{|l|l|l|}
\hline
$n$   & \qquad $1/\sqrt{\omega}$   &    $1/\sqrt{\omega}$ (from~\cite{Taut}) \\ \hline
2  &   0; $2.5198$  & 2 \\
3  &   0; 7.3825  & $4.4721$ \\
4 &   2.47047; 14.1004  & 2.2938; 7.3985 \\
5 &   0; 7.65075  & 4.97; 10.738  \\
\hline
\end{tabular}
}
\end{table}
\renewcommand{\arraystretch}{1}

We have also calculated the roots for $l=1$, as shown in Table~\ref{roots1}.
\newpage

\renewcommand{\arraystretch}{1.8}
\begin{table}[hbtp]
\caption{\small Some roots of $1/\sqrt{\omega}$ for $l =1$ and different values of $n$.}
\vspace*{0.2cm}
 \label{roots1}
\centerline{\small
\begin{tabular}{|l|l|}
\hline
$n$   & \qquad $1/\sqrt{\omega}$    \\ \hline
2  &   0; 3.6342  \\
3  &   0; 8.1091   \\
4 &   3.5028; 14.5564  \\
5 &   0; 8.3627; 21.6111  \\
\hline
\end{tabular}
}
\end{table}
\renewcommand{\arraystretch}{1}

Once the roots are determined, one can compute the energy levels and write dow, from Eq.~(\ref{ynl}) the explicit form of the functions $y_{nl}$ ($1\leq n \leq 5$). Some values for the low states are given in Table~\ref{energies}.

Similarly to Ref.~\cite{Taut_2010}, for $n=2, 3, 5$, we find solutions corresponding to the limit $\omega \rightarrow \infty$ ($1/\sqrt{\omega} \rightarrow 0$), which was called an asymptotic solution. Notice, however, that for $n=2$ such kind of solution was not reported in   Ref.~\cite{Taut_2010}.

\renewcommand{\arraystretch}{1.8}
\begin{table}[hbtp]
\caption{\small Comparison between our results and those of Refs.~\cite{Taut,Taut_94} for the lower energy values of different low $n$ states and for $l=0$. The first column presents the energy calculated by the Hellman-Feynman theorem (in 3D) \cite{Taut}. The second column displays the results of Ref.~\cite{Taut_94} and the third one displays our equivalent results, both in two-dimensions.}
\vspace*{0.2cm}
 \label{energies}
\centerline{\small
\begin{tabular}{|c|c|c|c|}
\hline
$n$   & $\epsilon^\prime$   &    $\epsilon_{int}^\prime$ & $\eta_{n0}$ \\ \hline
2  &   0.6250 & 1.000  & 0.4725 \\
3  &   0.1715 & 0.250  & 0.0734 \\
4 &    0.8553 & 0.108  & 0.8192 \\
5 &    0.2228 & 0.592  & 0.1025  \\
\hline
\end{tabular}
}
\end{table}
\renewcommand{\arraystretch}{1}

\newpage
The polynomials are given by:

\begin{eqnarray}
\label{ynl-geral}
 \nonumber y_{1l} &=& A_0 + \frac{A_1}{(2l+1)}\, r \\
 \nonumber \ \\
 \nonumber y_{2l} &=& y_{1l} + \frac{A_2 \sqrt{\omega}}{2(2l+1) [(2l+1)\sqrt{\omega}+1]}\, r^2 \\
  \nonumber \ \\
  \nonumber y_{3l} &=& y_{2l} + \frac{A_3 (\sqrt{\omega})^2}{6(2l+1) [(2l+1)\sqrt{\omega}+1][(2l+1)\sqrt{\omega}+2]}\, r^3 \\
  \nonumber \ \\
  \nonumber y_{4l} &=& y_{3l} + \frac{A_4 (\sqrt{\omega})^3}{24(2l+1) [(2l+1)\sqrt{\omega}+1]\ldots [(2l+1)\sqrt{\omega}+3]}\, r^4 \\
  \nonumber \ \\
   y_{5l} &=& y_{4l} + \frac{A_5 (\sqrt{\omega})^4}{120(2l+1) [(2l+1)\sqrt{\omega}+1] \ldots [(2l+1)\sqrt{\omega}+4]}\, r^5
\end{eqnarray}
where
{\small\begin{eqnarray}
\label{a_coef}
 \nonumber A_0 &=& 1 \\
 \nonumber \ \\
 \nonumber A_1 &=& - \left(\frac{1}{2\sqrt{\omega}}\right) \\
  \nonumber \ \\
  \nonumber A_2 &=& \left(\frac{1}{2\sqrt{\omega}}\right)^2 - 4 (2l+1) \sqrt{\omega} \\
  \nonumber \ \\
  A_3 &=& - \left(\frac{1}{2\sqrt{\omega}}\right)^3 + \frac{4}{\sqrt{\omega}} + 6 (2l+1) \\
  \nonumber \ \\
   \nonumber A_4 &=& \left(\frac{1}{2\sqrt{\omega}}\right)^4 - \frac{8}{(\sqrt{\omega})^2} - 6 (2l+1) \left[ \frac{1 }{\sqrt{\omega}} - 16 \sqrt{\omega} + 8 \right] \\
   \nonumber \ \\
  \nonumber A_5 &=& - \left(\frac{1}{2\sqrt{\omega}}\right)^5 + \frac{10}{(\sqrt{\omega})^3} - \frac{192}{\sqrt{\omega}} + (2l+1) \left[ \frac{5}{(\sqrt{\omega})^2} - 24 [1+4(2l+1)]\sqrt{\omega} - 400 \right]
\end{eqnarray}
}
Knowing the polynomials $y_{nl}$, the non-normalized wave functions $u_{nl}(r)$ are given by Eq.~(\ref{u_into_y}), namely
\begin{equation}\label{wavefunction_final}
 u_{nl} (r) = r^{l+1/2}\, e^{-\omega_n r^2/2}\, y_{nl}(r)
 \end{equation}
and
\begin{equation}\label{Rnl}
\psi(\vec r) = r^l\, e^{-\omega_n r^2/2}\, y_{nl}(r)\, e^{\pm il\theta} = R_{nl} (r,\omega_n) e^{\pm il\theta}
 \end{equation}
where each $\omega_n$ is a root of the respective $A_n$.

Let us now determine the wave function normalization. The highest polynomial we are considering here is of the type $y = 1 + Ar + Br^2 + Cr^3 + D r^4 + Er^5$. In such case, its square can be written as
$$ y^2 = \sum_{n=0}^{10} c_n r^n$$
where the coefficients $c_n$ are defined by: $c_0 = 1$, $c_1 = 2A$, $c_2 = A^2 + 2B$, $c_3 = 2(C + AB)$; $c_4 = B^2 + 2(D + AC)$; $c_5 = 2(E+AD+BC)$; $c_6= C^2+ 2(AE + BD)$; $c_7 = 2(BE +CD)$; $c_8 = D^2 + 2CE$; $c_9 = 2DE$ and $c_{10} = E^2$. Therefore, the normalization factor for this radial state is $N_{nl} = 1/\sqrt{I}$, where
$$I = \sum_{n=0}^{10} c_n\,  \int_0^\infty e^{-\omega r^2}\, r^{(2l+1 +n)}\, \mbox{d}r$$
All the integrals are of the type
$$\int_0^\infty \exp(-\mu r^p)\, r^{\nu -1}\, \mbox{d}r = \frac{1}{p}\, \mu^{-\nu/p}\, \Gamma\left(\frac{\nu}{p}\right)$$
So,
$$\displaystyle N_{nl} = \left[\frac{1}{2}\, \sum_{k=0}^{10} c_k \, \left(\frac{1}{\sqrt{\omega}}\right)^{2(l+1)+k}\, \Gamma \left(l+1+\frac{k}{2}\right)\right]^{-1/2}$$

For the normalization of the state $n=4$ we have just to put $E=0$ in the previous formula; for $n=3$, $E=D=0$ and so on. The normalization of each $(n,l)$ state is shown in Table~\ref{normalization}.

\renewcommand{\arraystretch}{1.8}
\begin{table}[hbtp]
\caption{\small Normalization factors for $n=2,3,4,5$ and $l=0,1$ states.}
\vspace*{0.2cm}
 \label{normalization}
\centerline{\small
\begin{tabular}{|c|c|}
\hline
$(n,l)$   & $N_{nl}$ \\ \hline
(2,0) & 0.240372  \\
(2,1) & 0.0486791  \\
(3,0) & 0.010507  \\
(3,1) & 0.00186614  \\
(4,0) & 0.00124923  \\
(4,1) & 0.0413502  \\
(5,0) & 0.017836  \\
(5,1) & 0,00139954  \\
\hline
\end{tabular}
}
\end{table}
\renewcommand{\arraystretch}{1}

Let us examine the system response for a microwave external excitation, which corresponds to the choice $\Omega = 0.06 - 0.6$~THz or 0.01~Ha $< \Omega <$~0.1~Ha (remember that $\omega = \Omega/2$).

A specific program was developed by the authors in $C++$ language and both calculations and graphics shown below were done by using the CERN/ROOT package.

\begin{figure}[htb!]
\centerline{\includegraphics[width=100mm]{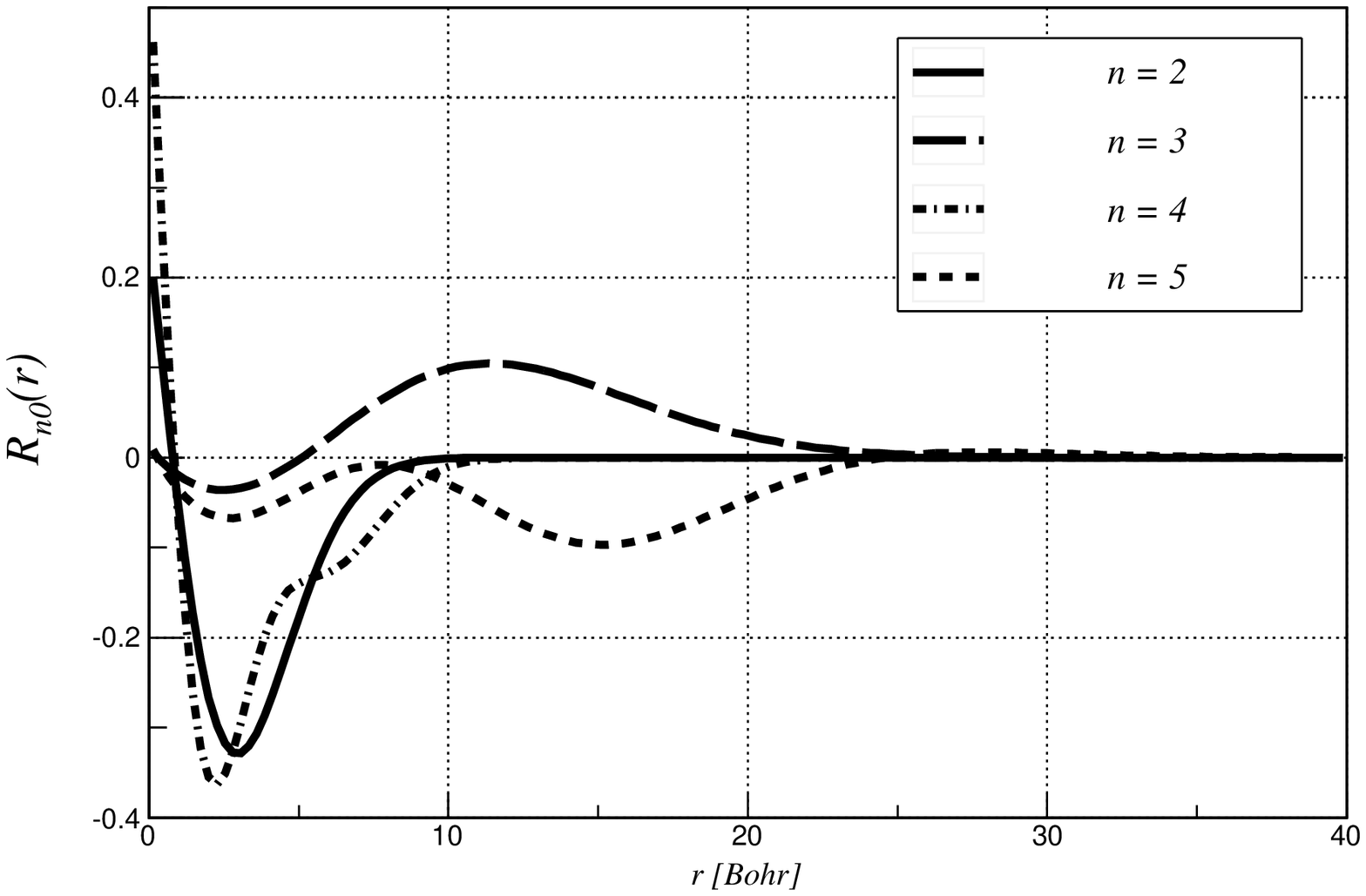}}
\caption{Radial wave function $R_{nl}$, for $l=1$ and $n=2,3,4,5$.}
\label{fig.1}
\end{figure}

\begin{figure}[htb!]
\centerline{\includegraphics[width=100mm]{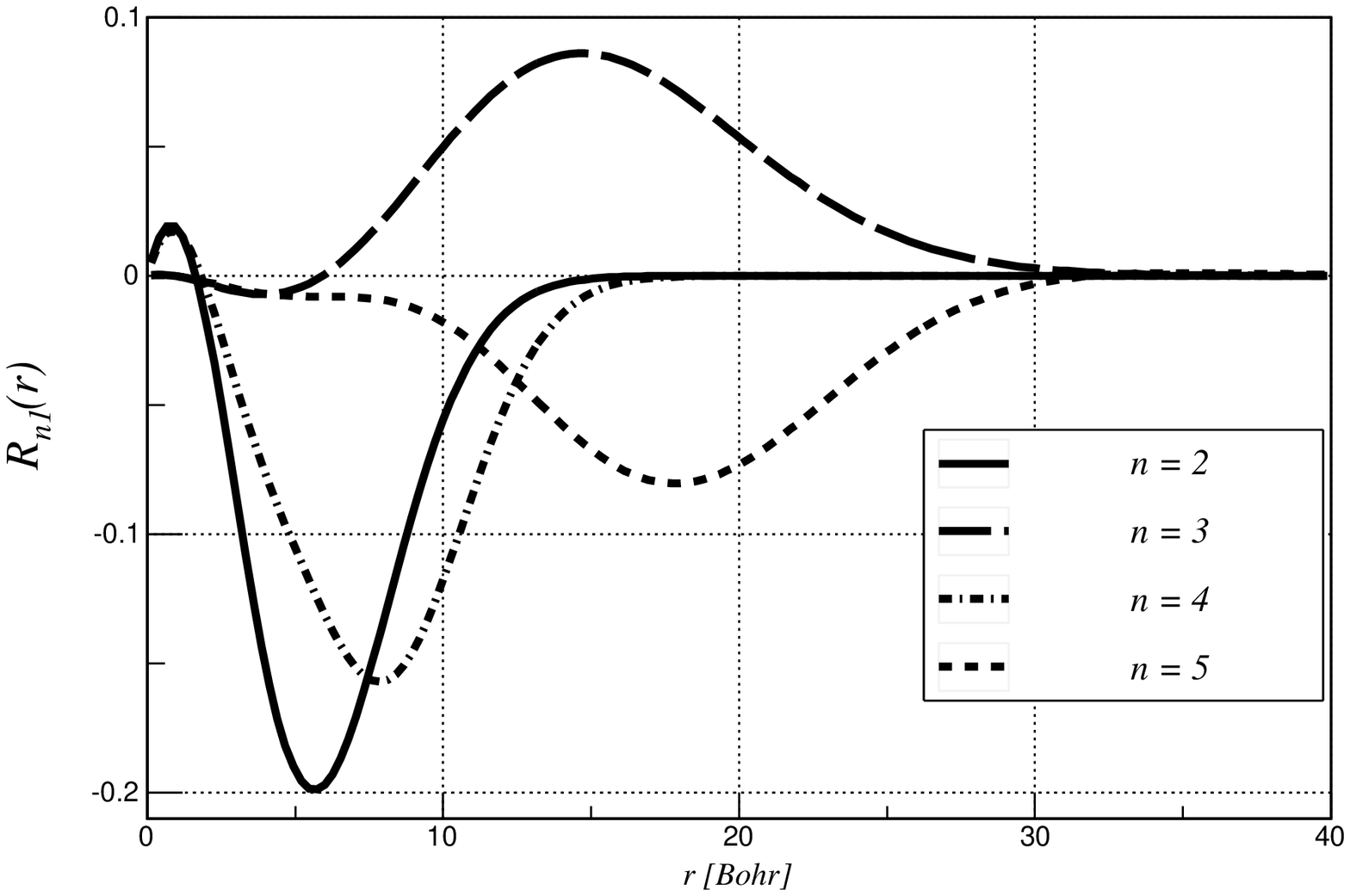}}
\caption{Radial wave function $R_{nl}$, for $l=1$ and $n=2,3,4,5$.}
\label{fig.2}
\end{figure}

Fig.~\ref{fig.1} and Fig.~\ref{fig.2} present a plot of how the radial wave function $R_{nl}$, defined by Eq.~(\ref{Rnl}), depends on $r$ and $\omega$, respectively, for $n=1$ and $n=5$ (both for the $s$ state). In the case $l=0$, the behavior of the states $n=2,3$ are similar to $n=1$ while the case $n=4$ is more like $n=5$. These graphics lead us to fix from now on the value of $\omega=0.01$~Ha in all the evaluations that follow.

\newpage
In the next figure we plot the wave functions $R_{40}(r)$ for two possible frequencies.

\begin{figure}[htb!]
\centerline{\includegraphics[width=100mm]{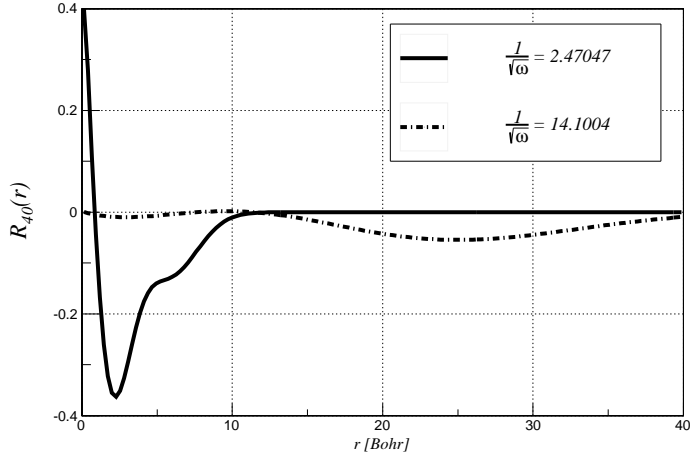}}
\caption{The behaviour of the radial wave function $R_{3l}$, for all the possible $l$ values.}
\label{fig.3}
\end{figure}

\begin{figure}[htb!]
\centerline{\includegraphics[width=90mm]{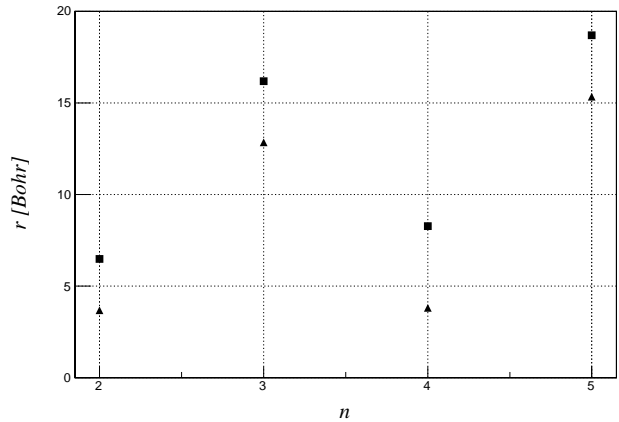}}
\caption{The mean value of the distance between the electrons, for different $n$ states. The legend of the figure is: triangle ($l=0$) and square ($l=1$).}
\label{fig.4}
\end{figure}

The mean values $<r>$ of the distance between the electrons were calculated for the $(n=2,3,4,5;l=0)$ states discussed in the paper, and are shown in Fig.~\ref{fig.4} and in Table~\ref{erre}.

\renewcommand{\arraystretch}{1.8}
\begin{table}[hbtp]
\caption{\small The mean values $<r>$ for $n=2,3,4,5$ and $l=0,1$ states.}
\vspace*{0.2cm}
 \label{erre}
\centerline{\small
\begin{tabular}{|c|c|}
\hline
$(n,l)$   & $\langle r_{nl} \rangle$ [Bohr] \\ \hline
(2,0) & 3.66754  \\
(2,1) & 6.48012  \\
(3,0) & 12.8597  \\
(3,1) & 16.1769  \\
(4,0) & 3.79963  \\
(4,1) & 8.26335  \\
(5,0) & 15.3347  \\
(5,1) & 18.6872  \\
\hline
\end{tabular}
}
\end{table}
\renewcommand{\arraystretch}{1}

\section{Discussions}\label{comm}

Let us start remembering that the approach followed in this paper allows us to get a quantization relation between the energy of the quantum dot and the external frequency, namely, $\eta_{nl} = (n + l + 1)\Omega/2$. Indeed, there is no need to recourse to the Hellman-Feynman theorem to calculate the energy corresponding to the frequencies for which analytical solutions have been found \cite{Taut}.

%It was numerically verified that for frequencies outside the range of microwave ($\Omega = 0.06 - 0.6$~THz or 0.01~Ha $< \Omega <$~0.1~Ha) there is no response to the external harmonic excitation.

So far the energy values are considered, we can see that there is a significant difference between our values and those found in Ref.~\cite{Taut_94} for all states from $n=2$ to $n=5$.

The smaller distance between the two electrons is found to be $<r> \simeq 3.7$~Bohr (for the state $n=2$ and $l=0$) (Fig.~\ref{fig.4}). The length scale of all the values shown in Table~\ref{erre} is compatible with the semiconductor lattice parameter as should be expected for quantum dots, \textit{i.e.}, electrons confined in a nanometer-scale semiconductor structure. Therefore, for the set of analytical solutions found here, all the configurations have a mean characteristic distance of the electrons in a quantum dot, $<r>$, comprehended in the range of 3.7 to 18.7~Bohr, in the case of external oscillations in the microwave frequency range.

We hope that the technique developed here can be use to find solutions of the Schr\"{o}dinger equation for other complex potential like, for example, the Kratzer potential which contains both a repulsive part and a long-range attraction and is relevant to describe some molecular systems, and for the charmonium and bottomonium states submitted to a confining phenomenological Cornell potential. Also the cases of quasi-exactly solvable potential for Klein-Gordon and the Dirac equations can be analyzed within the framework of the present paper.

We are now investigating numerical solution for quantum dots \cite{Caruso}, following the technique of Ref.~\cite{Caruso_2}, considering the $1/r$ and the $\ln r$ Coulombian potentials. 

\section*{Acknowledgment}
The authors are pleased to thank Bartolomeu D.B. de Figueiredo for useful suggestions and comments, and are indebt to an anonymous referee for pertinent and constructive criticism as well as for valuable comments.

%% References


\bibliographystyle{elsarticle-num}



\end{document}